\documentclass[final,5p,times,bibauthor,twocolumn]{elsarticle}
\usepackage{graphicx}
\usepackage{subfigure}
\usepackage{multirow}
\usepackage{multicol}
\usepackage{mathrsfs}
\usepackage[ps2pdf,bookmarks,colorlinks]{hyperref}

\journal{Nuclear Physics B}
\begin{document}
\begin{frontmatter}

\title{\bf Measurement of Cosmic Ray elemental composition from the CAKE 
balloon experiment}
\author[label1,label2]{S. Cecchini}
\author[label2]{T. Chiarusi}
\author[label2,label3]{G. Giacomelli}
\author[label2]{E. Medinaceli}
\author[label2]{L. Patrizii}
\author[label2]{G. Sirri}
\author[label2]{V. Togo}
\address[label1]{INAF-IASF, via P. Gobetti 101, 40129 Bologna, Italy}
\address[label2]{INFN Section of Bologna v.le C. Berti Pichat 6/2, Bologna,
  I-40127, Italy}
\address[label3]{Dept. of Physics, Univ. of Bologna, v.le C. Berti Pichat 6/2, Bologna,
  I-40127, Italy}
\begin{abstract}
CAKE (Cosmic Abundances below Knee Energies) was a prototype balloon
experiment for the determination of the charge spectra and of
abundances of the primary cosmic rays with nuclear charge Z$>$10. It was a passive
instrument made of layers of CR39$^{\textregistered}$ and 
Lexan$^{\textregistered}$/Makrofol$^{\textregistered}$ 
nuclear track detectors; it had a geometric acceptance of $\sim$0.7 m$^2$sr
for Fe nuclei. Here, the scanning and analysis strategies, the algorithms
used for the off-line filtering and for the tracking in automated mode of the
primary cosmic rays are presented, together with the resulting CR charge
distribution and their abundances.
\end{abstract}
\end{frontmatter}
\vspace{1cm}
\section{Introduction}\label{sec:intro}
The determination of the elemental abundances of cosmic rays (CRs) observed
in the solar system provides valuable information on the nature of their
sources, acceleration mechanisms and propagation in
the Interstellar Medium (ISM). While there is a general consensus for
supernova shocks being the main acceleration sites of CR particles with
energy per nucleon up to $10^{15}$ eV, still unresolved is the
question of which are their sources. 

From measurements of the abundances made near the Earth it is possible, by modelling the
propagation processes in the ISM, to infer the elemental abundances at CR sources as well
as to place constraints on the nature of their original environments.
The present data cannot select one model between coronal
material of later-type stars (which are rich of ions in a gas-phase) and Interstellar Medium
in the proximity of Supernovae Remnants.
Two proposed CR selection mechanisms based on the elemental First Ionization
Potential (FIP) and on the volatility/refractory state, \emph{i.e.} on the condensation 
temperature $T_c$, could justify  the first scenario with respect to the second. 
The measured CR abundances show a strong prevalence of the refractory
elements (which are generally characterized by low FIP and large $T_c$, e.g. Mg ($Z=12$),
 Ca ($Z=20$), Fe ($Z=26$) \cite{waddington})
with respect to the volatile ones (characterized by high FIP and small $T_c$,
e.g. H ($Z=1$), He ($Z=2$), O ($Z=8$)). The apparent anti-correlation between FIP and $T_c$
prevents to choose between the two mechanisms. Such degeneracy could be
solved by measuring the abundances of the few volatile low-FIP elements
and the rare refractory with high-FIP ones \cite{meyer}; e.g. Ga ($Z=31$), Ge ($Z=32$), 
Rb ($Z=37$), Pb ($Z=82$), Bi ($Z=83$) \cite{waddington}.

A number of signatures exists in the abundance
spectrum of heavy nuclei ($30<Z<74$) \cite{waddington}; in order
to reach confident conclusions it is necessary to achieve
high-resolution charge measurements and high statistics. Elements with
$Z>28$ are rare and large exposure factors (collecting area $\times$
exposure time $\times$ angular acceptance) are required. For example
the abundances of $Z>30$ nuclei is 1000$\div$10000 times smaller than Fe nuclei,
whose flux is $\sim$1 cm$^{-2}$sr$^{-1}$day$^{-1}$.

Since many years high altitude balloon flights provided
an alternative to more expensive (even in terms of
realization time) satellite based experiments. Flight durations have
been growing from few days to several weeks on recent Long Duration
Balloons (LDB) flown from Antartic regions  e.g. \cite{tiger}, and plans are 
underway to provide missions up to months (Ultra LDB).

Plastic passive detectors are promising since they can provide
good charge resolution, and are certainly suitable for balloon
flights, as they can be used in large collecting areas and
limited weight. The main limitation is that the detector has to be recovered
after the flight for the analysis with microscopes. Such type of technique
was already used succesfully in several experiments on 
balloons [4-6] 
and in 
space [7-10] 

Nuclear Track Detectors (NTDs) like CR39$^{\textregistered}$ can detect
charged particles with $Z/\beta\geq5$ with very good charge resolution 
\cite{bologna}, whereas Lexan$^{\textregistered}$/Makrofol$^{\textregistered}$ polycarbonates
 have higher thresholds,
 $Z/\beta\geq50$ \cite{makrofol0}. Due to the large flux of low charged particles the analysis
 of CR39$^{\textregistered}$ detectors and the determination of the relative abundances can
 be time consuming. Thus the necessity to
 develop an automatic system for reading and measuring the track
characteristics.

The purpose of the CAKE balloon experiment from the ASI 
(\emph{Agenzia Spaziale Italiana}) was
to test the implementation of new scanning and analysis techniques for
future LDB flights of large area detectors (tens of m$^2$sr) \cite{cake1}.
We report here about the scanning procedures and algorithms developed to analyse the
 CR39$^{\textregistered}$ 
sheets exposed during a trans-mediterranean flight.  The obtained primary charge 
spectrum at the top of the atmosphere is presented.
\vspace{-0.3cm}
\section{The detector}\label{sec:detector}
\begin{figure}[htb]
 \centering\resizebox*{6cm}{5cm}{\includegraphics{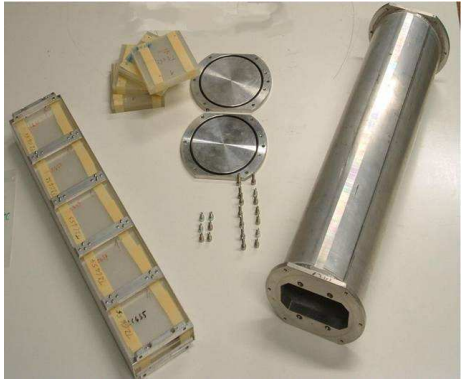}}\par
\caption{\small Composition of a cylindric aluminium container, and one
  assembled tray with 5 stacks.}\label{pic:cilinder}
\end{figure}

CAKE  was composed of 80 multilayer stacks of CR39$^{\textregistered}$ and Lexan$^{\textregistered}$ sheets. Each 
"standard'' 
stack was of dimension 11.5$\times$11.5 cm$^2$, with 10 sheets of Lexan$^{\textregistered}$, 0.25 mm thick,
sandwiched between 20 layers of CR39$^{\textregistered}$, 0.7 or 1.4 mm thick. Ten stacks were also 
assembled with Al or Pb foils
to study the fragmentation of the impinging particles. Each Al or Pb foil was 0.5 mm thick.
The thickness of a standard stack was $\sim$2.3 g/cm$^2$ while it was $\sim$ 3.0 
and 4.8 g/cm$^2$ for a stack with Al and Pb targets, respectively. Five stacks were 
inserted into an aluminium cylinder, as shown in Fig. \ref{pic:cilinder}, and four cylinders
were loadged
into aluminium ``boxes'' with an internal insulation foam coverage. The entire experimental
set-up composed of four ``boxes'' was lodged on the
payload of a stratospheric balloon. The cylinders and the boxes were designed to 
preserve the sheets in an air environment at normal preassure during the flight. 
The full exposed area was $\sim$1 m$^2$. 
  
\section{The test flight}\label{sec:flight}
The balloon was launched from the Trapani-Milo base (12.5$^{\circ}$E
32.92$^{\circ}$N) of ASI and landed in central Spain after 22 hours of 
flight.
The plafond altitude was 37$\div$40 km (3$\div$3.5 g cm$^{-2}$) and was
kept for about 17.8 hours. Along the balloon trajectory the mean
vertical rigidity cut-off was $\sim$7 GV (limiting the minimum energy
of the impinging CRs to E$>$2.6 GeV/nucleon) \cite{smart}. The gondola was not
azimuthally controlled and thus no East-West effect could be searched for.

During the balloon flight the cylinders had an internal preassure of 1 Atm,
and the inside temperature never exceeded
$34^{\circ}C$. Studies performed in a controlled heating room,
determined that CR39$^{\textregistered}$ keeps its sensitivity unchanged from
$-50^{\circ}C$ up to $50^{\circ}C$ \cite{tom}. 
\section{Data Analysis}\label{sec:scan}
\begin{figure}[htb]
\centering\resizebox*{7cm}{6cm}{\includegraphics{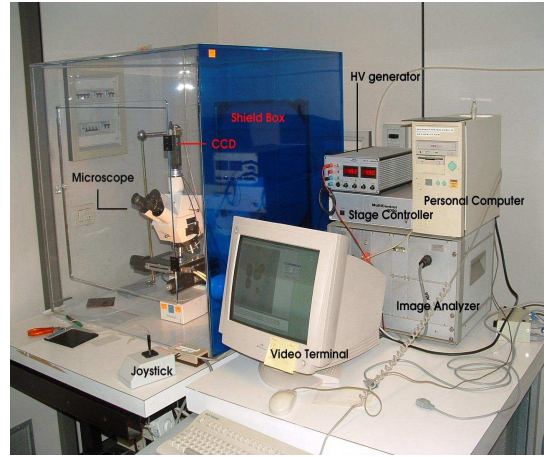}}\par
\caption{\small The used automatic measuring system.}\label{pic:elbek}
\end{figure}

After the flight CR39$^{\textregistered}$ sheets of standard stacks 
were etched in a 6N NaOH solution at
(70$\pm$0.1)$^{\circ}$C for 30 hours; the total number of detector
foils that were chemically etched was $\sim$280. The detectors were measured by
an automated SAMAICA (Scanning and Measuring with Automatic Image
Contour Analysis) image analyzer, \emph{i.e.} an integrated
DAQ system composed of an optical microscope controlled by a
personal computer, a built in frame grabber, and a positioning stage
\cite{sam}; the image analyzer was a product of ELBEK-Bildanalyse GmbH;
see Fig. \ref{pic:elbek}.
\vspace{0.2cm}

Chemical etching of NTDs results in the formation of conical pits (etch-pits) on both
sides of a detector foil. The surface base of the etch-pits has an elliptical shape.

For each exposed stack four CR39$^{\textregistered}$ detector plates were measured in
automatic mode. For each etch-pit (``track'') 
on the upper surface of a CR39$^{\textregistered}$ plate the system provides the measurement of
the major and minor axis, polar angle, area of the base of each cone, plus its position
on the foil. The scanned area of each foil was reduced to 8$\times$8 cm$^2$ to avoid
edge effects; a single detector sheet was measured in $\sim$12 hours. 

By determining the CR39$^{\textregistered}$ bulk etching rate $v_B$ \cite{makrofol} it was possible to
compute the reduced etching rate $p=v_T/v_B$, where $v_T$ is the etching velocity
along the particle trajectory. Exposures of CR39$^{\textregistered}$ stacks to relativistic heavy ion beams
at accelerators \cite{calibCR39} allowed a correlation of the measured $p$
with the charge number $Z$ of the impinging ions \cite{durrani}.
We had at our disposal three CR39$^{\textregistered}$ calibrations for three different $v_B$ values, 
namely 1.1, 1.15, and 1.21 $\mu m/h$ \cite{cake2}. The interval
 $1.05~\mu m/h\leq v_B\leq 1.25~\mu m/h$ was used for accepting the bulk velocity 
of the etched foils. 236 out of 280 detectors were selected and analyzed.
The parameters for the automatic measurement of the detectors were set to 
obtain a minimum bias in the track measurement; the residual noise was treated
off line with the implementation of an appropiated filter. The scan efficiency was
$>~$90\% for the detection of tracks from nuclei with charges
$Z/\beta\geq$ 9 \cite{tom}. A filter was implemented in order to classify the
measured tracks using a multilayer perceptron neural network \cite{perceptron} 
trained on a data sample of $\sim$3000 cosmic ray tracks measured
manually. The filter selection is based on two features of
the measured tracks: the track ellipticity (defined as the
ratio of the minor to major semi axes of the base etch-pit) and its central
brightness (a quantity measured in arbitrary units by the optical system).
The filtered data set are $\sim~$60\% of the raw data set \cite{lalo}.

Semi-automatic scans were made by operators to check the goodness of
the automatic measurements and the filter; for each track it was verified if it was
passing through both sides of the detector foil.
The automatic scan is $\sim$10 times faster than the semi-automatic scan.
\vspace{0.6cm}
\begin{figure}[htb]
\centering\resizebox*{6.0cm}{3.0cm}{\includegraphics{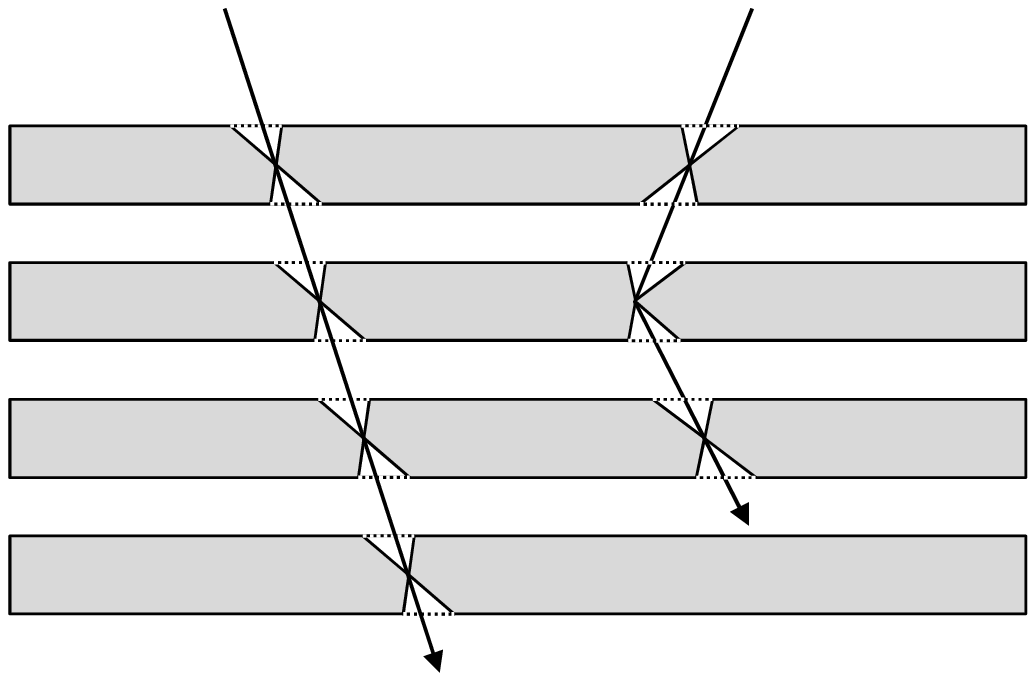}}\par
\vspace{-0.2cm}
\caption{\small A stack composed of 4 CR39$^{\textregistered}$ sheets. Charged primary CRs
  pass through the stack with linear trajectories, like the trajectory at the left.
  The tracks in consecutive detectors are matched and
  linked recursively. Selected tracks are those passing through $\geq3$ detectors. Tracks
  with non linear trajectories, as the trajectory on the right, 
  are rejected}\label{pic:3of4}
\end{figure}

\begin{figure}[htb]
\centering\resizebox*{6cm}{6cm}{\includegraphics{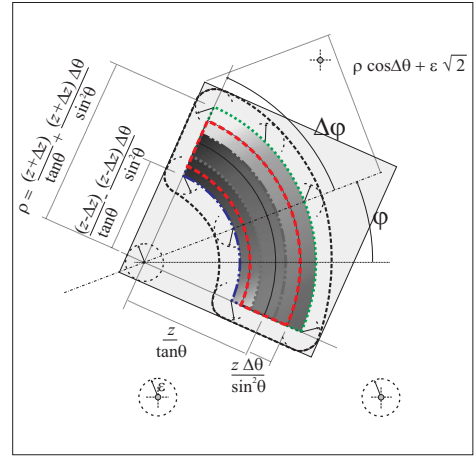}}\par
\caption{\small ``Fiducial'' area where a candidate CR track is
searched for in subsequent foils. The area takes into account the experimental 
uncertainties in zenith ($90^{\circ}-\theta$), azimuth $\varphi$, the distance between
consecutive foils $z$, and the mis-alignment beween them $\epsilon$.}\label{pic:area}
\end{figure}

The trajectory of each candidate ion was followed through 3 CR39$^{\textregistered}$ sheets;
Fig. \ref{pic:3of4} shows different tracks crossing several foils of the
same stack. Using tracking algorithms only tracks crossing the plates
in linear trajectory were selected (cosmic ray particles with energies $E\geq3$
GeV/nucleon are highly penetrating and scattering is negligible). A cosmic ray
``event'' was defined as a track passing through at least three consecutive
detector plates as shown by the trajectory on the left in Fig. \ref{pic:3of4}.
Each candidate event was tracked individually by means of the 
\emph{fiducial area recursive method} \cite{cakepub}. Starting with a ``track''
on the uppermost foil of the stack the method defines a limited area on the 
following foils where corresponding tracks are searched for. This procedure is
applied recursively up to 3 etched foils on the stack. The fiducial area
sketched in Fig. \ref{pic:area} was determined considering the experimental
uncertainties of the track inclination (in Fig. \ref{pic:area} $\theta$ is the 
complementary of the zenith angle) and its orientation ($\varphi$ is the azimuth angle)
with respect to the foil uppermost surface. All the uncertainties of the 
measured variables are considered. \emph{i.e.} the etch-pit base
area ($\pm$10 $\mu$m$^2$), the foil thickness (at most $\pm$50 $\mu$m)
and the mis-alignment among the foils in the stack. The size of the
fiducial area was $<$1 $cm^2$. Details about the tracking
and selection procedures are given in ref. \cite{tom,lalo}.
\vspace{-0.1cm}
\vspace{-0.2cm}
\section{Results}\label{sec:results}
\begin{figure}[htb]
\centering\resizebox*{8.85cm}{7.85cm}{\includegraphics{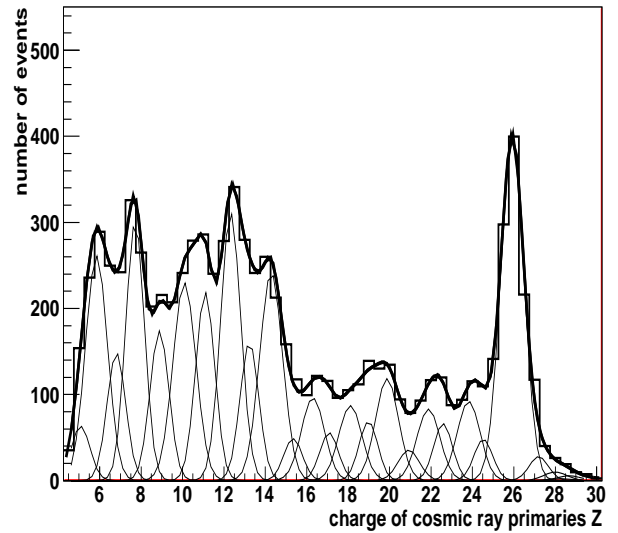}}\par
\caption{\small Experimental charge distribution (histogram). A global fit (thick black curve)
 is applied using a Gaussian approximation for the individual peaks (represented 
 by the thin curves).}\label{pic:fitx}
\end{figure}

In Fig. \ref{pic:fitx} is shown the unfolded charge distribution of the CR events
detected in CR39$^{\textregistered}$. The range of the reconstructed charge number Z is $5\leq Z \leq 30$.
In the same figure are shown the Gaussian distributions of each element (represented with 
thin black curves) which all added together fits the experimental histogram (represented with
the thick black curve). The Gaussian parameters (height and width) were computed by means of a
global fit of the sum of 25 Gaussians to the experimental distribution. The fit was computed
using the least squares minimization package \cite{minuit}. The standard deviations of the
fitted curves were in the range 0.4 $\div$ 0.6 Z. By integrating over each individual
Gaussian distribution we obtained the distribution of observed events at the detector 
level shown in Fig. \ref{pic:int}.

\vspace{-0.2cm}
\vspace{-0.2cm}
\subsection{Fragmentation processes}\label{subse:frac}
\begin{figure}[htb]
\vspace{-0.3cm}
\centering\resizebox*{8.1cm}{7.2cm}{\includegraphics{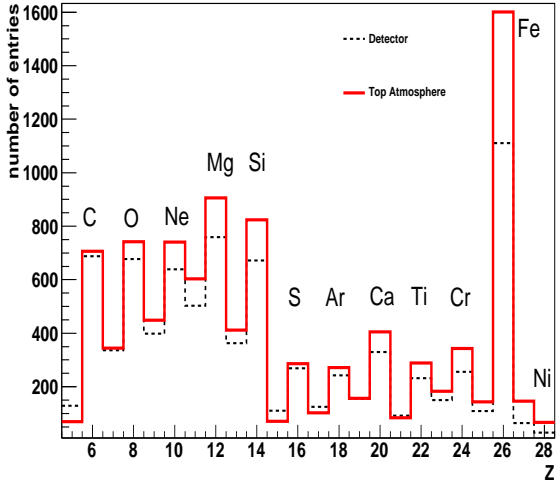}}\par
\vspace{-0.3cm}
\caption{\small Charge distribution of the unfolded distribution with a
  resolution of 1$e$ in the range $5\leq Z \leq 30$. The dotted black histogram
  is the distribution of events for each species at the detector level. The solid red
  histogram is the charge distribution at the top of the atmosphere obtained by correcting for 
  fragmentation processes in the aluminium containers
  of the detectors and in the residual atmosphere.}\label{pic:int}
\end{figure}

In order to compute the CR charge distribution at the top of the
atmosphere, fragmentation processes have to be considered \cite{frag}. 
Fragmentation of cosmic ray nuclei may occur in the interactions with the residual 
atmosphere; the mean atmospheric thickness at the CAKE trajectory was 3.25 $g/cm^2$. 
Fragmentation also occurs in the interaction with the aluminium 
containers housing the detector stacks, shown in Fig. \ref{pic:cilinder};
the mean Al thickness was 2.7 $g/cm^2$. Fragmentation on air and on Al were modelled with 
Monte Carlo simulations using Geant4 \cite{geant}. The model takes into account the
interactions of CR nuclei with the residual atmosphere and the detector shields.
Primary CR energies were sampled assuming the energy dependence of the flux $\Phi\cong E^{-2.7}$ 
\cite{wiebel-sooth} in the range $5\leq Z \leq 28$. With the simulations the fragmentation 
coefficients of heavy elements into lighter species were obtained, calculated as in ref.
\cite{orth} and \cite{ichimura};
then with the fragmentation matrix the charge distribution at the top of the atmosphere
was obtained. In Fig. \ref{pic:int} the black histogram shows the
charge distribution at the detector level; the solid red histogram is the charge 
distribution at the top of the atmosphere.
\vspace{-0.3cm}
\subsection{Angular Acceptance}\label{subsec:acc}
\begin{figure}[h!]
\centering\resizebox*{7.75cm}{6.75cm}{\includegraphics{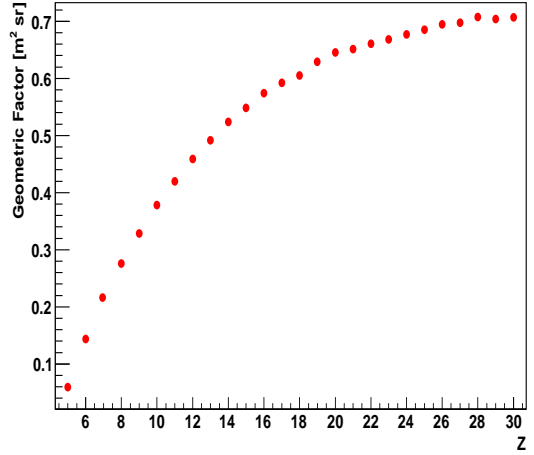}}\par
\vspace{-0.3cm}
\caption{\small Angular acceptance multiplied by the exposure area and the
  exponential attenuation factors as a function of the
  charge $Z$.}\label{pic:acc}
\end{figure}

The geometrical acceptance of the C39 detectors varies with the charge Z \cite{durrani} 
as there is a limiting angle $\theta_C$ for the formation and observation of a track
in the detectors
\begin{equation}
\sin(\theta_C)=\frac{1}{p} = \frac{v_B}{v_T}
\end{equation}
By including the exponential attenuation of 
CRs (i.e. by considering the residual atmosphere, the Al containers of the detectors
and the insulating foam which filled the vessels) the geometrical factor of the
detector varies as shown in Fig. \ref{pic:acc}.
\vspace{-0.3cm}  
\subsection{Relative Abundances at the top of the Atmosphere}
\begin{figure}[h!]
\centering\resizebox*{8.0cm}{7.0cm}{\includegraphics{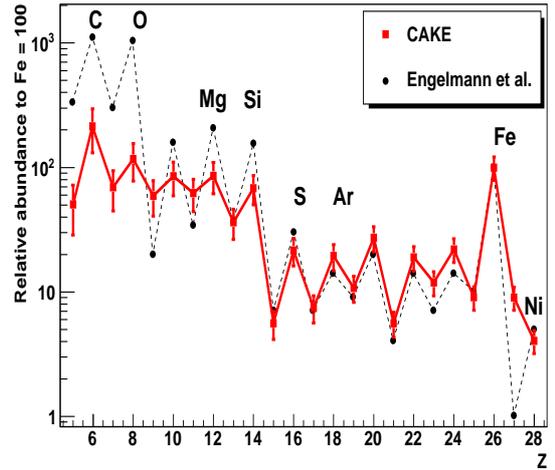}}\par
\caption{\small Measured CAKE abundances relative to iron (set equal to 100), as a function
 of the charge (square points and red solid line). Even charge elements are quoted in the plot.
Error bars correspond to systematic and statistical standard deviations added in quadrature.
Experimental measurements of a similar experiment 
compiled in ref. \cite{engelmann} are also plotted as black points interpoled by a 
dashed black line.}\label{pic:finCAKE}
\end{figure}

The abundances of nuclei at the top of the atmosphere, were evaluated considering 
the angular acceptance and correcting for the flux attenuation in air and in the 
detector boxes. The resulting values are shown in Fig. \ref{pic:finCAKE} 
together with the measurement reported in ref. \cite{engelmann} at a similar
energy threshold (2.6 GeV/nucleon). The agreement is satisfactory for $12\leq Z\leq 26$.
For $Z<12$ there is a disagreement due to the filter and algorithm rejecting small 
tracks affecting mainly low charged particles.
The differences at the left of the Fe peak are probably due to a partially unsatisfactory
correction for fragmentation. At the right of the Fe peak the number of events is
too small to even consider a significant result.

In Fig.$~$\ref{pic:finCAKE} error bars in CAKE's data represent systematic plus statistical 
standard deviations added in quadrature; the error around the Iron relative abundance 
is $\sim$ 20\%.
\vspace{-0.2cm}
\section{Conclusions}\label{sec:conclusions}
This paper was focused on the development of automatic analysis procedures
for determining the chemical composition of cosmic ray nuclei using nuclear track
detectors. The NTD technique offers good charge resolution and is 
attractive for the exploration/determination of the charge spectrum of cosmic particles
at reasonable cost. The explotation of this technique in large area 
experiments relies on automatic image analyzers for scanning and
measuring the detector foils as well as on the availability of LDB and ULDB flights.

By making use of a short stratospheric balloon flight we had the 
opportunity to study and develop several tools to obtain a charge spectrum of
cosmic ray particles with $Z/\beta\geq10$ at the top of the atmosphere. Our measurement is 
in fair agreement with other observations. 
\vspace{-0.3cm}
\section*{Acknowledgements}
We acknowledge the support of the Trapani-Milo Stratospheric Balloon Launch Base of ASI, 
and the cooperation of the technical staff at the INFN Section of Bologna. The experiment
was funded by ASI. One of the authors thanks INFN for a post-doc fellowship for 
foreigners.

\end{document}